# Resource-efficient domain adaptive pre-training for medical images

## Authors

### Yasar Mehmood

**Affiliation 1:** Department of Computer Science, COMSATS University Islamabad, Lahore Campus. Pakistan (http://lahore.comsats.edu.pk/default.aspx)

**Affiliation 2:** Department of Computer Science & IT, Virtual University of Pakistan, Lahore. Pakistan (https://www.vu.edu.pk/)

**ORCID:** 0000-0002-2891-9129

### *Usama Ijaz Bajwa (Corresponding Author)

**Email Address:** usamabajwa@cuilahore.edu.pk

**Affiliation:** Department of Computer Science, COMSATS University Islamabad, Lahore Campus. Pakistan (http://lahore.comsats.edu.pk/default.aspx)

**Full postal address:** COMSATS University Islamabad, Lahore Campus, 1.5 KM Defence Road, Off Raiwind Road, Lahore Pakistan

**ORCID:** 0000-0001-5755-1194

### Xianfang Sun

**Email Address:** SunX2@cardiff.ac.uk

**Affiliation:** School of Computer Science and Informatics, Cardiff University, Abacws, Senghennydd Road, Cardiff, CF24 4AG. UK. (https://www.cardiff.ac.uk/computer-science)

**ORCID:** 0000-0002-6114-0766

# Resource-efficient domain adaptive pre-training for medical images

## Abstract

The deep learning-based analysis of medical images suffers from data scarcity because of high annotation costs and privacy concerns. Researchers in this domain have used transfer learning to avoid overfitting when using complex architectures. However, the domain differences between pre-training and downstream data hamper the performance of the downstream task. Some recent studies have successfully used domain-adaptive pre-training (DAPT) to address this issue. In DAPT, models are initialized with the generic dataset pre-trained weights, and further pre-training is performed using a moderately sized in-domain dataset (medical images). Although this technique achieved good results for the downstream tasks in terms of accuracy and robustness, it is computationally expensive even when the datasets for DAPT are moderately sized. These compute-intensive techniques and models impact the environment negatively and create an uneven playing field for researchers with limited resources. This study proposed computationally efficient DAPT without compromising the downstream accuracy and robustness. This study proposes three techniques for this purpose, where the first (partial DAPT) performs DAPT on a subset of layers. The second one adopts a hybrid strategy (hybrid DAPT) by performing partial DAPT for a few epochs and then full DAPT for the remaining epochs. The third technique performs DAPT on simplified variants of the base architecture. The results showed that compared to the standard DAPT (full DAPT), the hybrid DAPT technique achieved better performance on the development and external datasets. In contrast, simplified architectures (after DAPT) achieved the best robustness while achieving modest performance on the development dataset[1].

## 1. Introduction

The data scarcity for deep learning-based analysis of medical images is a well-known problem. Two of the main reasons are the high cost to annotate the data and the patients' privacy. For example, it took almost an hour of radiologists' time to label one patient in the brain tumor segmentation (BraTS) dataset (Menze et al., 2014). Similarly, a challenge has been launched recently to generate synthetic medical data that preserves patients' privacy used for training the models (Jordon et al., 2021). The method for transfer learning, where the weights trained on a pre-training task are used as initial weights for the target task, has been used traditionally for data-scarce domains. The recent architectures are relatively deeper and contain a lot of trainable parameters, and insufficient training data quickly leads to overfitting. For transfer learning, the independent and identical distribution (IID) is not needed for training data which allows the architecture to pre-train on a large dataset from a different domain and then transfer this knowledge to the target domain with scarce data (Tan et al., 2018).

Several open-source pre-trained models are available for researchers to use in their data-scarce target tasks. Most of these models are pre-trained on generic datasets like ImageNet (Deng et al., 2009), and many studies have used ImageNet pre-trained models for medical image analysis tasks (Haghighi, Taher, Zhou, Gotway, & Liang, 2021; Shin et al., 2016; Tajbakhsh et al., 2016; Y. Yang et al., 2018). However, it has been investigated that transfer learning with the domain differences between the pre-training and downstream data is not very useful (Azizi et al., 2021; Hosseinzadeh Taher, Haghighi, Feng, Gotway, & Liang, 2021; Raghu, Zhang, Kleinberg, & Bengio, 2019; Reed et al., 2022). Other studies highlighted the better results achieved using the in-domain data during pre-training (Chen, Ma, & Zheng, 2019; Geyer, Corinzia, & Wegmayr, 2019; Heker & Greenspan, 2020; G. Liang & Zheng, 2020; Taleb et al., 2020).

---

[1] Code to reproduce the results is available at the GitHub repository https://github.com/mehmoodyasar/Resource-Efficient-Domain-Adaptive-Pre-Training.git

Recent techniques have addressed this problem by pre-training using data progressively similar to the target domain after initializing the model from the generic dataset pre-trained weights. For example, the study (Azizi et al., 2021) initialized the model with the ImageNet pre-trained weights and then pre-trained the model using the in-domain data. Similarly, the studies (Hosseinzadeh Taher et al., 2021; Reed et al., 2022) used the data progressively similar to the target domain and improved performance. These studies compared the results of this domain adaptive pre-training (DAPT) with the generic dataset pre-training and in-domain dataset pre-training to conclude that the DAPT (also called hierarchical pre-training) was the best technique in terms of accuracy for the downstream tasks. Also, DAPT proved to be more robust to distribution shifts, which is necessary for a deep learning-based system to be deployed in the real world.

The dataset needed for DAPT is much smaller than generic datasets like ImageNet, but it should be significantly larger than the target dataset. Moreover, the studies (Azizi et al., 2021; Hosseinzadeh Taher et al., 2021; Reed et al., 2022) performed DAPT on an entire architecture (like ResNet50). DAPT is computationally expensive compared to transferring the knowledge from ImageNet pre-trained weights, where typically, a few top layers are fine-tuned on the target task. The report (Bommasani et al., 2021) highlighted that foundational models would become significant contributors to global carbon emissions with their ever-increasing sizes. In addition to the negative environmental impact, the compute-intensive techniques hamper the ability of the low-resourced researchers to conduct research and benefit from the latest techniques to achieve SOTA performance.

The proposed study aims to make the DAPT computationally efficient by taking inspiration from a study that conducted a thorough investigation on the use of the ImageNet pre-trained models for medical image analysis (Raghu et al., 2019). The authors concluded that most of the benefit of transferring knowledge is limited to the base architecture's first few layers/blocks. This behavior was attributed to the over-parameterized pre-trained architectures and the domain differences between the pre-training and downstream data. Therefore, the study proposed simpler variants of the base architecture, including the first few layers/blocks of the base architecture and a simple architecture on top of it. Another inspiration was taken from the standard transfer learning method as recommended by François Chollet in the Keras blog (Chollet, 2020). According to this method, fine-tuning may be performed only on a few top layers of the architecture instead of fine-tuning all the layers using the downstream data. A study (Tajbakhsh et al., 2016) showed that fine-tuning a sufficient number of layers can achieve better results than training all the layers using pre-trained weights. This study proposes three classes of techniques to make the DAPT efficient. DAPT is performed only on a subset of layers (partial DAPT) instead of the complete architecture in the first class. In the second class (called hybrid DAPT), partial DAPT is performed for a few epochs, and then for the remaining epochs, DAPT is performed on all the layers (full DAPT). Finally, simplified variants of the base architecture are created in the third class, and the corresponding generic dataset pre-trained weights are transfused to these simplified architectures as initial weights before DAPT. Simplified variants improve the training efficiency and save resources during inference time. Although many studies have shown the superiority of the learned representations through self-supervised pre-training over the supervised pre-training (Azizi et al., 2021; Hosseinzadeh Taher et al., 2021), the study (Reed et al., 2022) showed that the benefits obtained through self-supervised DAPT hold even when supervised DAPT is used. Also, self-supervised learning is computationally expensive and involves lengthy calculations. This study, therefore, uses supervised DAPT while making it efficient, accurate, and robust.

The metric of the area under the curve (AUC) on the downstream classification was used to evaluate the quality of the representations learned from the proposed techniques. The learned model on the downstream task was evaluated on an external dataset to evaluate the robustness of these representations. The performance of different strategies on the development and external dataset was compared in relation to the computational resources needed (the size of the architecture for storage needs and the energy required to train the network (Lannelongue, Grealey, & Inouye, 2021; T.-J. Yang, Chen, Emer, & Sze, 2017)).

The contributions and findings of this study are:

- ✓ Exploration of the trade-off between accuracy on the target task and computational efficiency of the DAPT
- ✓ Exploration of the proposed efficient DAPT techniques' impact on the robustness of the models trained for the downstream tasks
- ✓ The representations learned through partial DAPT (which is a computationally efficient technique than full DAPT), when performed on a sufficient number of layers, can outperform the full DAPT on the downstream task (development dataset). However, full DAPT performs better when this downstream model is evaluated on an external cohort.
- ✓ The representations learned through the proposed hybrid strategy achieve better accuracy and robustness for the downstream tasks than the representations learned through partial DAPT or full DAPT.
- ✓ The representations learned through simpler variants of the base architecture (ResNet50) are the most robust while achieving modest accuracy.

## 2. Related Work

### 2.1 Transfer Learning in Medical Imaging Domain

The most common form of transfer learning is when the weights trained on a surrogate task are used to initialize a model on some downstream task. It has been used for data-scarce domains like medical imaging in deep architectures that require many samples to avoid overfitting. This technique allows the data-scarce domains to use complex architectures and makes the training process computationally efficient by reusing the weights of the pre-trained model. Therefore, one-time pre-training can be reused multiple times across many downstream tasks. Many studies used transfer learning using ImageNet pre-trained weights for the downstream tasks involving medical images (Haghighi et al., 2021; Shin et al., 2016; Tajbakhsh et al., 2016; Y. Yang et al., 2018). However, the domain differences between the pre-training data and downstream data limit the efficacy of this type of transfer learning. Studies showed limited benefits of using ImageNet weights for the downstream tasks of medical image analysis (Azizi et al., 2021; Hosseinzadeh Taher et al., 2021; Reed et al., 2022). The study (Raghu et al., 2019) analyzed the performance of transferring learning from generic dataset pre-training to the downstream tasks involving medical images and highlighted the critical role of the domain difference in hampering the downstream accuracy. Other studies also showed better results when pre-training and downstream task data were from the same domain (Chen et al., 2019; Geyer et al., 2019; Heker & Greenspan, 2020; G. Liang & Zheng, 2020; Taleb et al., 2020).

**2.2 Bridging the Domain Differences through DAPT**

Recent studies have proposed an efficient (computationally and data-wise) technique for addressing the problem of domain difference between pre-training and downstream task data (Azizi et al., 2021; Hosseinzadeh Taher et al., 2021; Reed et al., 2022). DAPT presented a promising approach for improving the accuracy of transfer learning in the downstream tasks by leveraging the open-source models pre-trained on generic datasets. Instead of using in-domain pre-training from scratch, researchers initialized the models using the generic dataset (e.g., ImageNet) pre-trained weights. Further pre-training was performed using the in-domain (or closer to the target domain) data. The studies concluded that DAPT outperformed models pre-trained on generic datasets and pre-trained on in-domain data alone.

**2.3 Computational Efficiency at Training and Inference Time**

The report (Bommasani et al., 2021) highlighted the importance of computational efficiency to improve the inclusiveness for those with limited resources and the positive role it will likely play in reducing the carbon footprints of the deep learning models both during training and inference. Many studies have been published to measure the environmental impact of their algorithms. For example, the study (Lannelongue et al., 2021) proposed a standardized tool for estimating the carbon footprint of any algorithm to provide a metric for the researchers to evaluate and improve their algorithms. Similarly, the study (Parcollet & Ravanelli, 2021) highlighted that the architectures that outperform SOTA by a tiny margin often have a much greater carbon footprint. The study (García-Martín, Rodrigues, Riley, & Grahn, 2019) presented a survey of the techniques to compute the computational cost in terms of the energy consumed by machine/deep learning models to raise awareness among the researchers for considering the computational cost of architectures while developing new techniques.

Other studies have proposed methods to achieve computational efficiency while maintaining an acceptable level of performance. For example, the study (Patterson et al., 2021) proposed efficient architectures to reduce the harmful environmental impact. On the other hand, the study by Google (Patterson et al., 2022) presented the best practices for machine/deep learning researchers to reduce carbon emissions. The study (T. Liang, Glossner, Wang, Shi, & Zhang, 2021) presented a survey of the network pruning and quantization techniques to make the architectures energy efficient. A critical study in this regard proposed a technique called lottery-ticket that was able to find sub-architectures (as small as up to 10% of the original network) within the complex architectures that achieve the accuracy comparable with the complete architecture (Frankle & Carbin, 2018). Finally, a recent study published in the Neural Information Processing (NeurIPS 21) conference proposed a robust and accurate technique for performing model compression (Diffenderfer, Bartoldson, Chaganti, Zhang, & Kailkhura, 2021). This study showed the superiority of lottery-ticket-based methods to preserve the accuracy while improving robustness with the compressed architectures. Different methods were also combined to achieve even better accuracy and robustness while being computationally efficient.

## 3. Research Methodology

This study has initialized the models with the ImageNet pre-trained weights, and DAPT has been performed using the BraTS 2020 dataset with the surrogate task of binary classification into low-grade glioma (LGG) vs. high-grade glioma (HGG). The extracted features after DAPT are then fed to the classifiers for downstream classifications of brain tumor type and brain disease. Also, the downstream classifier trained for brain tumor type classification has been evaluated on an external dataset to check the robustness of the features obtained after DAPT.

### 3.1 DAPT Settings

In this sub-section, all the strategies for DAPT have been explained. The convolution base of the ResNet50 architecture (He, Zhang, Ren, & Sun, 2016) has been used in the proposed study like the studies (Azizi et al., 2021; Hosseinzadeh Taher et al., 2021; Raghu et al., 2019; Reed et al., 2022). Internal details of the ResNet50 architecture and its various portions used in this study have been shown in Fig. A1 (Appendix A). Fig. A2 (Appendix A) shows the inner details of the different sub-blocks of the architecture. For all the strategies performing DAPT, the convolution base of ResNet50 (or its first one or two block(s)) is followed by the global average pooling and sigmoid layers because of the surrogate task of binary classification (LGG vs. HGG).

Firstly, the strategies which are based on the complete convolution base of the ResNet50 architecture are described. The first strategy named ImageNet is the one in which no DAPT is performed, and ImageNet weights are used as it is and is, therefore, least expensive computationally among the strategies using the complete ResNet50 architecture. The second one is the domain adapted (DA) strategy, in which all the layers are pre-trained domain adaptively, and it is, therefore, the full DAPT strategy. The subsequent strategies perform DAPT only on the last one sub-block (DA_L1SB) and the last two sub-blocks (DA_L2SB) of ResNet50 while keeping the rest of the layers frozen. These are the partial DAPT strategies where the architecture is partially pre-trained using the in-domain data.

In the hybrid DAPT strategies, also called the progressive fine-tuning (PFT) strategies, we perform partial DAPT using the last one and two sub-blocks for a few epochs, and then for the remaining epochs, we perform DAPT on all the layers. These strategies are called DA_L1SB_PFT (perform partial DAPT on the last one sub-block and then on all the layers) and DA_L2SB_PFT (perform partial DAPT on the last two sub-blocks and then on all the layers).

The third class of the strategies is based on simplified variants of the ResNet50 architecture. The first strategy is based on the first block of the ResNet50 (DA_TF_F1B), followed by the global average pool and sigmoid layers (Fig. A1). The second one is based on the first two blocks of ResNet50 (DA_TF_F2B), followed by the global average pool and sigmoid layers. The full DAPT is performed in both the strategies after initializing them with the ImageNet weights. The corresponding strategies, ImageNet_TF_F1B and ImageNet_TF_2B, also use the same architectures as DA_TF_F1B and DA_TF_F2B, respectively, but no DAPT is performed in them.

ImageNet_F1B is the cheapest of all the strategies because its architecture is the simplest, and no DAPT is performed. At the same time, ImageNet_TF_F2B and ImageNet are the second and third cheapest strategies, respectively.

Following the recommendation given in the Keras for transfer learning and fine-tuning (Chollet, 2020), for all the strategies performing DAPT (whether full, partial, or hybrid), all the layers in the convolution base are frozen, and only the sigmoid layer is trained for 1000 epochs. This step is performed to prevent the sigmoid layer, which is initialized using random weights, from destroying the ImageNet weights in the convolution base. It has been named phase 1 of DAPT. For strategies DA, DA_TF_F1B, and DA_TF_F2B, which perform full DAPT (domain adaptively pre-train all the layers), all the layers of the architecture are trained for the following 150 epochs, and this has been named as phase 2. Similarly, for the strategies, DA_L1SB and DA_L2SB, only the last one and two sub-blocks respectively are trained (keeping the other layers frozen) for the subsequent 150 epochs during phase 2. For the strategies DA_L1SB_PFT and DA_L2SB_PFT, partial DAPT is performed for 100 epochs during phase 2, while all the layers are trained for the following 50 epochs during phase 3.

All the strategies, along with the total and trainable parameters for different phases, have been shown in Table 1.

**Table 1.** Strategies used for domain adaptive pre-training (DAPT)

| Strategy Name | Architecture | Total Parameters | Phase | Trainable Parameters | Epochs | Energy Consumption in Kilowatt-Hour (kWh) |
|---|---|---|---|---|---|---|
| ImageNet | ResNet50 |  | - | - | - | - |
| DA | ResNet50 | 23,589,761 | 1 | 2,049 | 1000 | 94.86 |
|  |  |  | 2 | 23,483,521 | 150 |  |
| DA_L1SB | ResNet50 | 23,589,761 | 1 | 2,049 | 1000 | 67.93 |
|  |  |  | 2 | 4,461,569 | 150 |  |
| DA_L2SB | ResNet50 | 23,589,761 | 1 | 2,049 | 1000 | 70.12 |
|  |  |  | 2 | 8,921,089 | 150 |  |
| DA_L1SB_PFT | ResNet50 | 23,589,761 | 1 | 2,049 | 1000 | 76.90 |
|  |  |  | 2 | 4,461,569 | 100 |  |
|  |  |  | 3 | 23,483,521 | 50 |  |
| DA_L2SB_PFT | ResNet50 | 23,589,761 | 1 | 2,049 | 1000 | 78.64 |
|  |  |  | 2 | 8,921,089 | 100 |  |
|  |  |  | 3 | 23,483,521 | 50 |  |
| DA_TF_F1B | First One Block of ResNet50 | 230,017 | 1 | 257 | 1000 | 30.41 |
|  |  |  | 2 | 224,129 | 150 |  |
| DA_TF_F2B | First Two Block of ResNet50 | 1,460,609 | 1 | 513 | 1000 | 56.28 |
|  |  |  | 2 | 1,440,385 | 150 |  |
| ImageNet_TF_F1B | First One Block of ResNet50 | 229,760 | - | - | - | - |
| ImageNet_TF_F2B | First Two Blocks of ResNet50 | 1,460,096 | - | - | - | - |

### 3.2 Architectures for Downstream Classification

To check the quality of learned representations of the different DAPT strategies used, features were extracted after DAPT (or directly from the ImageNet weights in the strategies ImageNet, ImageNet_TF_F1B, and ImageNet_TF_F2B) and used in the following two downstream tasks:

- Brain disease classification
- Brain tumor type classification

The robustness of the model learned for the brain tumor type classification was evaluated on an external dataset. The downstream classification was performed using the two fully connected neural networks. The architectures of the two fully connected neural networks can be seen in Tables 2 and 3. It can be seen that model 1 is simpler while model 2 is relatively complex. The rationale for using the two models is to see the results when classifiers of varying complexity are used.

**Table 2.** Model 1 architecture

| Layer Number | Units | Activation |
|---|---|---|
| Layer 1 | 64 | ReLU |
| Layer 2 | 32 | ReLU |
| Layer 3 | 3 for brain tumor type classification / 5 for brain disease classification | Softmax |

**Table 3.** Model 2 architecture

| Layer Number | Units | Activation |
|---|---|---|
| Layer 1 | 512 | ReLU |
| Layer 2 | 256 | ReLU |
| Layer 3 | 256 | ReLU |
| Layer 4 | 128 | ReLU |
| Layer 5 | 3 for brain tumor type classification / 5 for brain disease classification | Softmax |

### 3.3 Evaluation Measures

For evaluating the performance of the representations learned by different DAPT strategies, the area under the curve (AUC) on the downstream classification tasks (development and external datasets) has been used like the studies (Azizi et al., 2021; Hosseinzadeh Taher et al., 2021; Reed et al., 2022). In this regard, strategies have been compared based on their overall AUC and how quickly they reached a threshold value of 0.9.

Two measures have been used to quantify the computational resources needed for each strategy. The first one is the total number of parameters in each strategy to quantify the memory needed. The second one is the energy consumption, considering the total running time of the strategy and the hardware configuration used. The freely available tool has been utilized to estimate energy consumption (Lannelongue et al., 2021). This direct energy estimation tool has been adopted because the number of parameters alone does not consistently measure the energy consumed by a particular neural network (T.-J. Yang et al., 2017). The estimate of energy consumption for each strategy in kWh has been given in Table 1. The exact method of estimating the energy consumption has been discussed in Appendix B.

## 4. Experimental Settings

### 4.1 Datasets and Tasks

The proposed study used the BraTS 2020 dataset (Bakas et al., 2017; Bakas et al., 2018; Menze et al., 2014) for DAPT, which contains magnetic resonance imaging (MRI) scans. The dataset contains LGG and HGG classes, and the surrogate task in DAPT is to perform binary classification of HGG vs. LGG. The detail of BraTS 2020 has been given in Table 4, while the example cases are shown in Fig. 1. For DAPT, slices containing at least one tumor pixel have been extracted from T1 weighted sequences. Only one sequence was used during the DAPT, despite the availability of four sequences, to test the transfer of knowledge across different sequences/modalities.

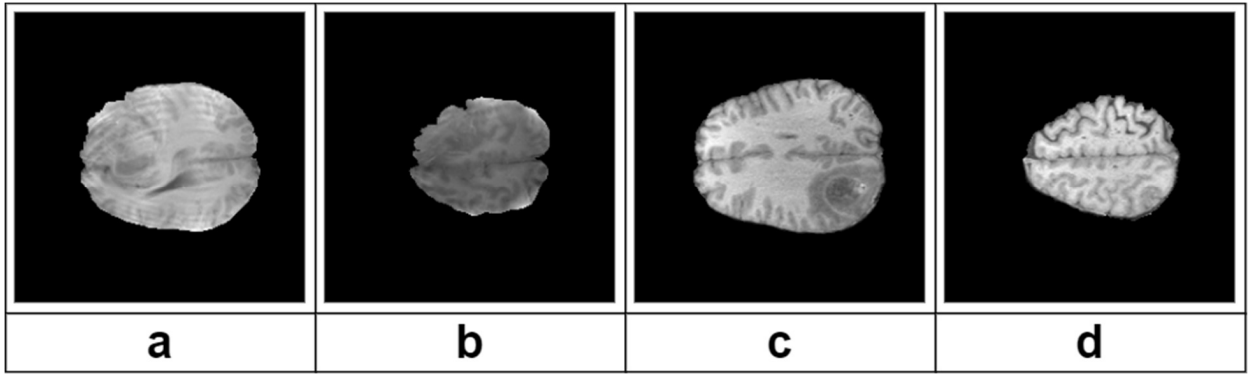

**Fig. 1.** Different slices/images from the BraTS 2020 dataset. **a.** and **b.** are of class LGG, while **c.** and **d.** are of class HGG. The images **a.** vs. **b.** and **c.** vs. **d.** show the difference in the sizes of brain masks at different locations within a 3D scan

**Table 4.** BraTS 2020 dataset used for DAPT

| Class | Imaging Modality | Slices/Images |
|---|---|---|
| LGG | T1 weighted | 4926 |
| HGG | T1 weighted | 19496 |

The target tasks are classification of brain diseases using Harvard Medical School (HMS) dataset (re3data.org, 2021) and brain tumor type classification using contrast-enhanced MRI (CE-MRI) dataset (Cheng, 2017). The dataset details for HMS and CE-MRI are shown in Tables 5 and 6, respectively, whereas Fig. 2 and 3 show a few examples of the HMS and CE-MRI datasets, respectively.

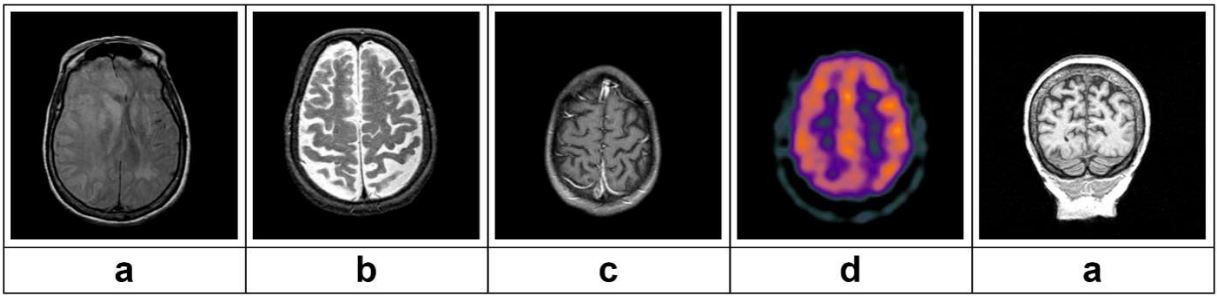

**Fig. 2.** One example each belonging to the five classes of the HMS dataset. **a.** image belonging to cerebrovascular class has the modality proton-density (PD) weighted, **b.** image belonging to degenerative class with the modality T2 weighted, **c.** image of inflammatory class with Gadolinium contrast-enhanced (GAD) modality, **d.** image of neoplastic class with single-photon emission computed tomography (SPECT) modality, and **e.** image of normal class with T1 weighted modality.

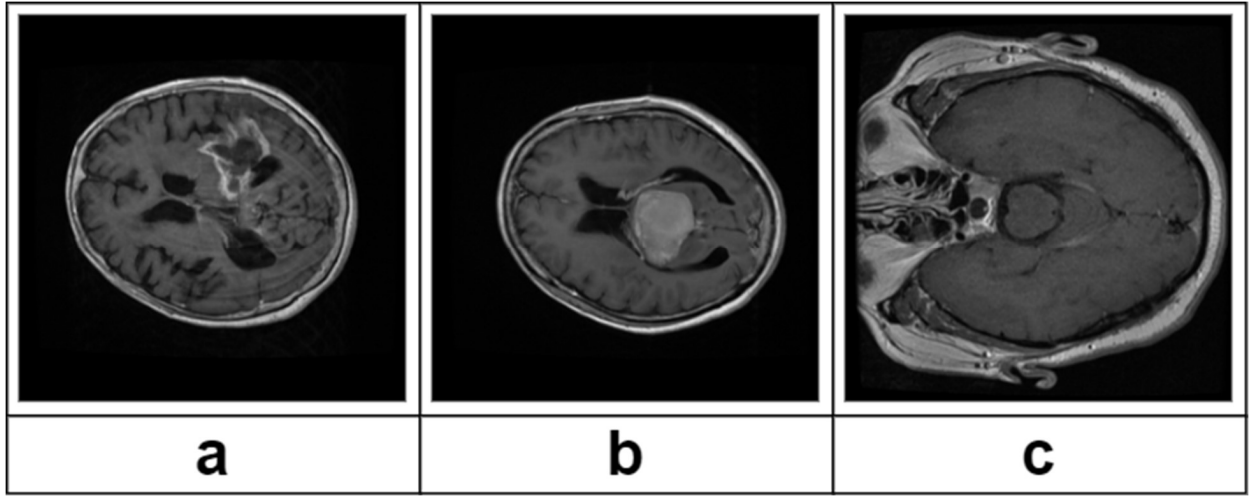

**Fig. 3.** One example each of the three classes of CE-MRI dataset. **a.** glioma, **b.** meningioma, and **c.** pituitary

**Table 5.** HMS dataset used for the downstream task of brain disease classification

| Class | Imaging Modalities | Slices/Images |
|---|---|---|
| Cerebrovascular | T1 Weighted, T2 Weighted, MR GAD, PD Weighted | 689 |
| Degenerative | T1 Weighted, T2 Weighted, PD Weighted, MR PD | 370 |
| Inflammatory | T1 Weighted, T2 Weighted, PD Weighted, MR GAD, | 463 |
| Neoplastic | T1 Weighted, T2 Weighted, PD Weighted, CT, MR GAD, SPECT, | 868 |
| Normal | T1 Weighted, T2 Weighted, PD Weighted | 314 |

**Table 6.** CE-MRI dataset used for the downstream task of brain tumor type classification

| Class | Imaging Modality | Cases |
|---|---|---|
| Meningioma | T1 weighted contrast enhanced | 708 |
| Glioma | T1 weighted contrast enhanced | 1426 |
| Pituitary | T1 weighted contrast enhanced | 930 |

Finally, the Kaggle brain tumor type classification (Kaggle BTTC) dataset (Bhuvaji, Kadam, Bhumkar, Dedge, & Kanchan, 2020) was used as an external dataset to evaluate the robustness of the proposed techniques for the models trained on the downstream task of brain tumor type classification. The dataset detail of Kaggle BTTC has been shown in Table 7, and the diversity of images in the dataset can be seen in Fig. 4.

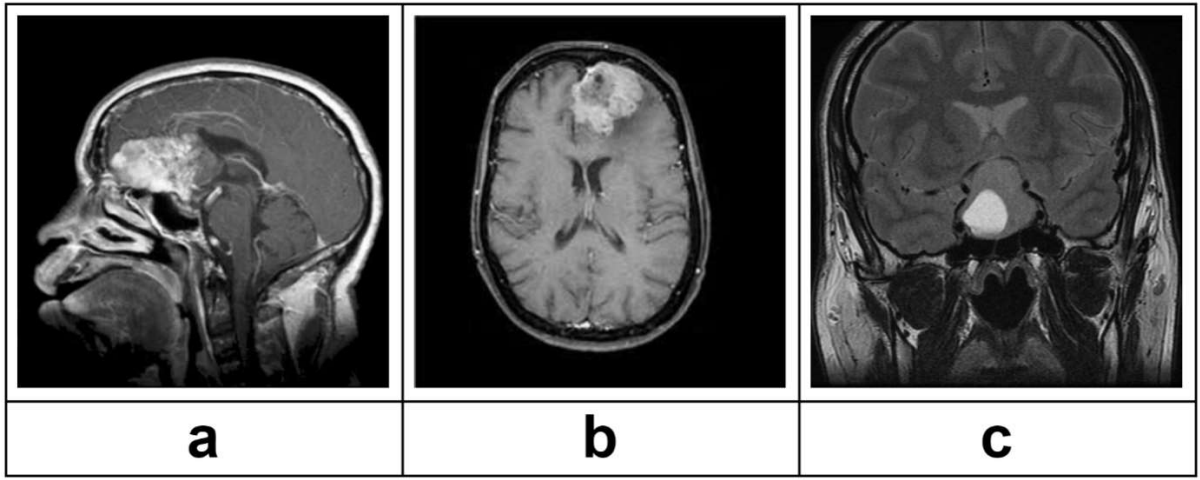

**Fig. 4.** One example each of the three classes of the Kaggle BTTC dataset. **a.** glioma, **b.** meningioma, and **c.** pituitary, where images are taken from different orientations

**Table 7.** Kaggle BTTC dataset used to check the robustness of the model trained on CE-MRI dataset

| Class | Imaging Modality | Cases |
|---|---|---|
| Meningioma | Unspecified | 937 |
| Glioma | Unspecified | 926 |
| Pituitary | Unspecified | 901 |

### 4.2 Implementation Details

For all the DAPT experiments, the optimizer Adam was used. The learning rate was 0.0001 for all the phase 1 training and 0.00001 for phase 2 and 3 training. Cloud-based computational resources using Google Colab Pro were used for the DAPT with the setting of the tensor processing unit (TPU) and high-RAM (~35 GB). The training data (BraTS 2020) was stored in NumPy arrays of float16 (2 bytes) instead of using float32 or float64 to save memory at the expense of higher precision calculations. The batch size for all the DAPT experiments was set to 16. Python language with Keras API was used for the implementation.

The class-balanced train-test-split of 0.8 : 0.2 was used for the downstream classification. Therefore, for brain tumor type classification, the train and test sets consisted of 2451 and 613 examples, respectively. Similarly the train and test sets consisted of 2163 and 541 examples for brain disease classification. All the experiments for downstream classification were conducted using Google Colab, which has around 12 GB RAM, and GPU was used as a hardware accelerator. For both the models, a learning rate of 0.00005 and Adam optimizer was used. Each model was trained for 500 epochs in downstream classification experiments. Like the DAPT experiments, Python language with Keras API was used for implementation.

## 5. Results and Discussion

In this section, the results of experiments are discussed on each downstream task (brain disease classification (Fig. 5 and 6) and brain tumor type classification (Fig. 7 and 8)), and robustness (Fig. 9 and 10). The performance of different strategies in relation to their computational cost (memory in terms of parameters and energy consumption in kWh) has been visualized in Fig. B1 (Appendix B). The strategies

ImageNet_TF_F1B and ImageNet_TF_F2B performed poorly for all the experiments and have not been included in all the Figures (except for Fig. 13, 14 and 15). The prior studies (Azizi et al., 2021; Hosseinzadeh Taher et al., 2021; Reed et al., 2022) performed DAPT by fine-tuning all the layers of the base architecture like the DA strategy of this research. Therefore, better results (accuracy, robustness, and computational efficiency) of the proposed DAPT strategies than the DA strategy represent the improvements of the proposed techniques over the state-of-the-art. The study's main findings have been summarized in the sub-section 5.4.

## 5.1 Brain Disease Classification

The results (test set AUC) of the brain disease classification have been shown in Fig. 5 and 6 for the downstream models 1 and 2, respectively. An important thing to keep in mind while viewing the results for brain disease classification is that the downstream data for this task is relatively more diverse and different from the pre-training data, as it includes CT images and sequences like SPECT (Fig. 2). As shown in Fig. 5, the strategy DA_L2SB_PFT reached the threshold (0.9 test AUC) first. This strategy (purple line) produced the best results for most epochs. The corresponding partial DAPT strategy DA_L2SB (red line) surpassed DA_L2SB_PFT for a few epochs (between 0 and 100), but overall, the hybrid strategy outperformed its corresponding partial DAPT strategy. The second best strategy for model 1 was DA_L1SB_PFT, a progressive fine-tuning (hybrid DAPT) technique, and it outperformed its partial DAPT counterpart DA_L1SB by a vast margin. The strategy DA was well below the partial and hybrid DAPT strategies. For model 2, DA_L2SB_PFT was again the winner by reaching the threshold test set AUC in only in seven epochs. Its simple counterpart DA_L2SB, and the DA_L1SB_PFT strategy were very close together in the second and third position to reach the threshold. However, the overall performance of DA_L2SB was better than DA_L1SB_PFT. For models 1 and 2, the hybrid DAPT strategies (with partial DAPT performed on enough layers in its phase 2) outperformed the corresponding partial and full DAPT (DA) strategies. The simplified architecture DA_TF_F1B achieved the least performance for both models. In contrast, the architecture DA_TF_F2B performed better for model 1 and outperformed DA_L1SB and ImageNet strategies while achieving comparable results as DA.

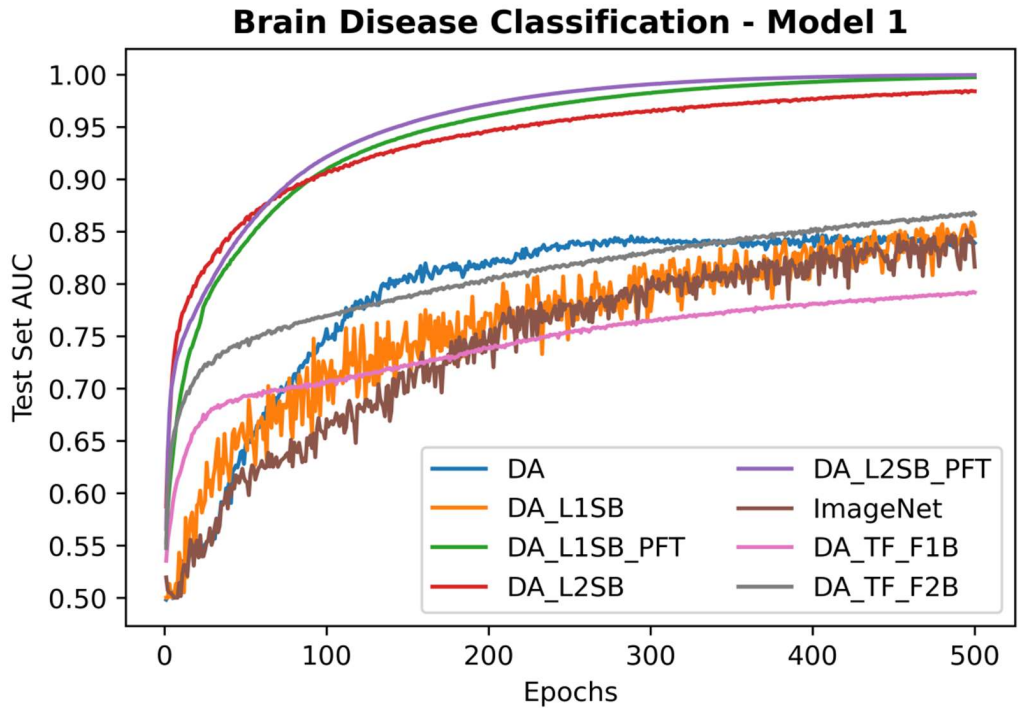

**Fig. 5.** The test set AUC for brain disease classification using HMS dataset for model 1

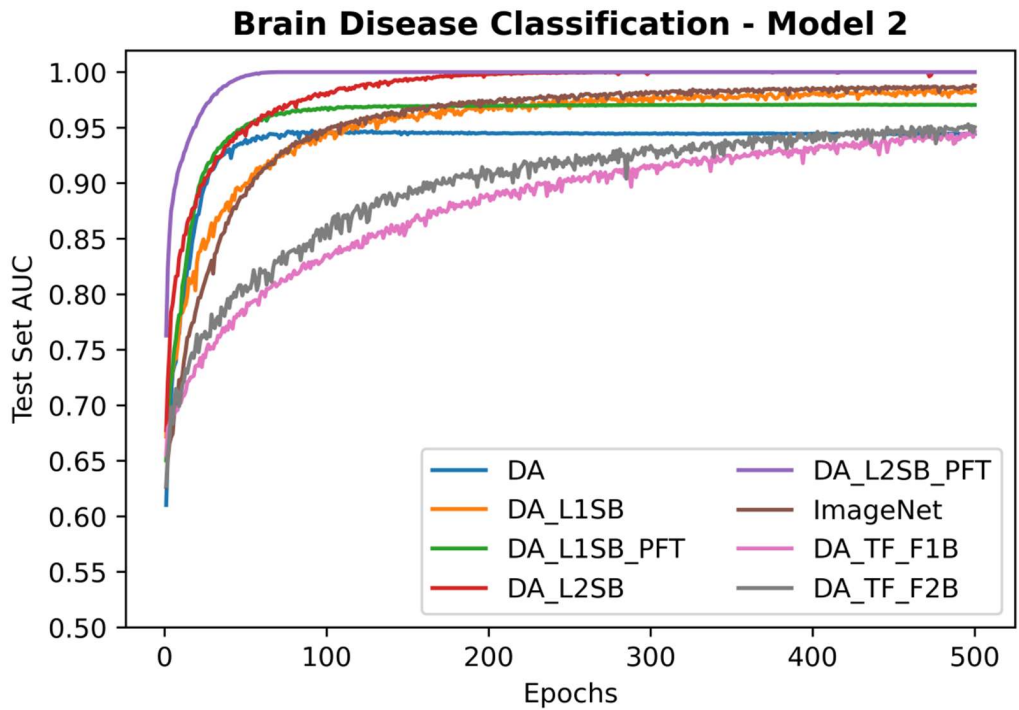

**Fig. 6.** The test set AUC for brain disease classification using HMS dataset for model 2

**5.2 Brain Tumor Type Classification**

The test set AUC results of brain tumor type classification using the CE-MRI dataset are shown in Fig. 7 and 8. Again, the hybrid DAPT strategies performed better, especially for model 2. For model 1, the strategy DA and DA_L2SB_PFT achieved comparable results while DA was the first (by a small margin) to reach the threshold sooner and achieve slightly better overall performance. The hybrid strategy of DA_L1SB_PFT (which performs DAPT on the last one sub-block during phase 2) was at the third position, followed by the DA_L2SB_PFT strategy's corresponding partial DAPT strategy (DA_L2SB). For model 2, DA_L2SB_PFT was the first to reach 0.9 AUC, followed by DA. DA_L2SB and simplified architecture DA_TF_F2B were the third and fourth to reach the threshold. It can be seen from Fig. 7 and 8 that the simplified architecture DA_TF_F2B outperformed ImageNet for both the models, DA_L1SB for model 1, and was at par with this strategy for model 2. The strategy DA_TF_F1B was again the most diminutive performer for both models.

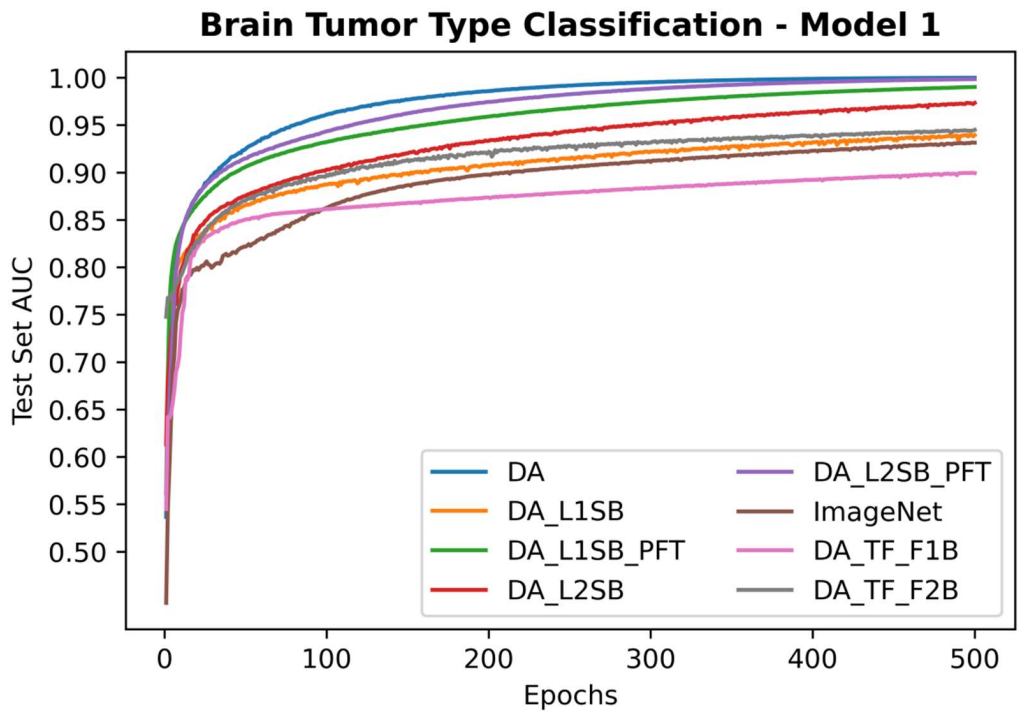

**Fig. 7.** The test set AUC for brain tumor type classification using CE-MRI dataset for model 1

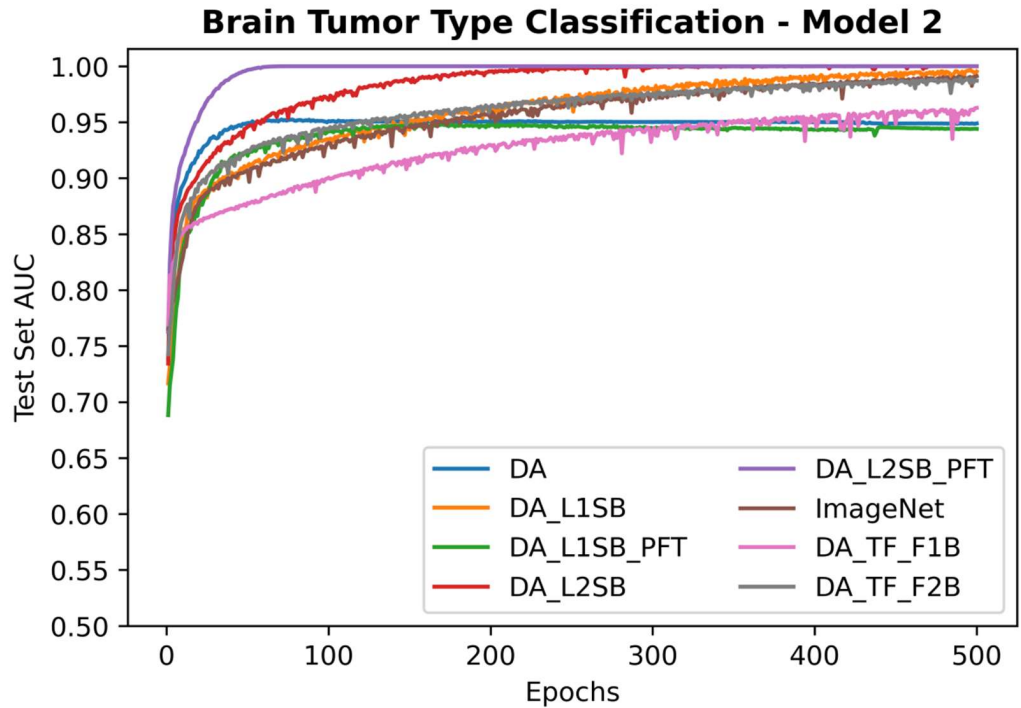

**Fig. 8.** The test set AUC for brain tumor type classification using CE-MRI dataset for model 2

### 5.3 Robustness

The models trained on the CE-MRI dataset were evaluated on the Kaggle BTTC dataset to check the robustness of all the strategies. Fig. 9 and 10 present the robustness experiments performed for models 1 and 2, respectively. Interestingly, the simpler architectures DA_TF_F1B and DA_TF_F2B outperformed all the strategies based on the complete ResNet50 architecture. Although none of the strategies crossed the threshold of 0.9 AUC, the best results were obtained with DA_TF_F1B followed by DA_TF_F2B, highlighting the simpler architectures' generalization ability. Both the strategies also maintained their performance levels (except DA_TF_F1B showing a slight dip after reaching the epoch close to 100 for model 2) from initial to later epochs compared to the other strategies based on the complete ResNet50 architecture. Different strategies of the complete ResNet50 architecture, which achieved outstanding results on the development dataset, performed poorly on the external cohort. Another significant finding is that the performance of hybrid DAPT techniques was very close to the full DAPT strategy of DA. In contrast, hybrid DAPT strategies performed better than the full DAPT strategy on the development datasets. In model 1, the overall performance of DA_L2SB_PFT was slightly better than DA, while the performance of DA_L1SB_PFT was slightly below it. Similarly, for model 2, DA_L2SB_PFT again performed slightly better than DA, while DA_L1SB_PFT was well below DA for most epochs (although the maximum AUC of DA_L1SB_PFT was higher than that of the DA strategy). The most significant dip in AUC from the earlier to later epochs was for the ImageNet strategy in model 2.

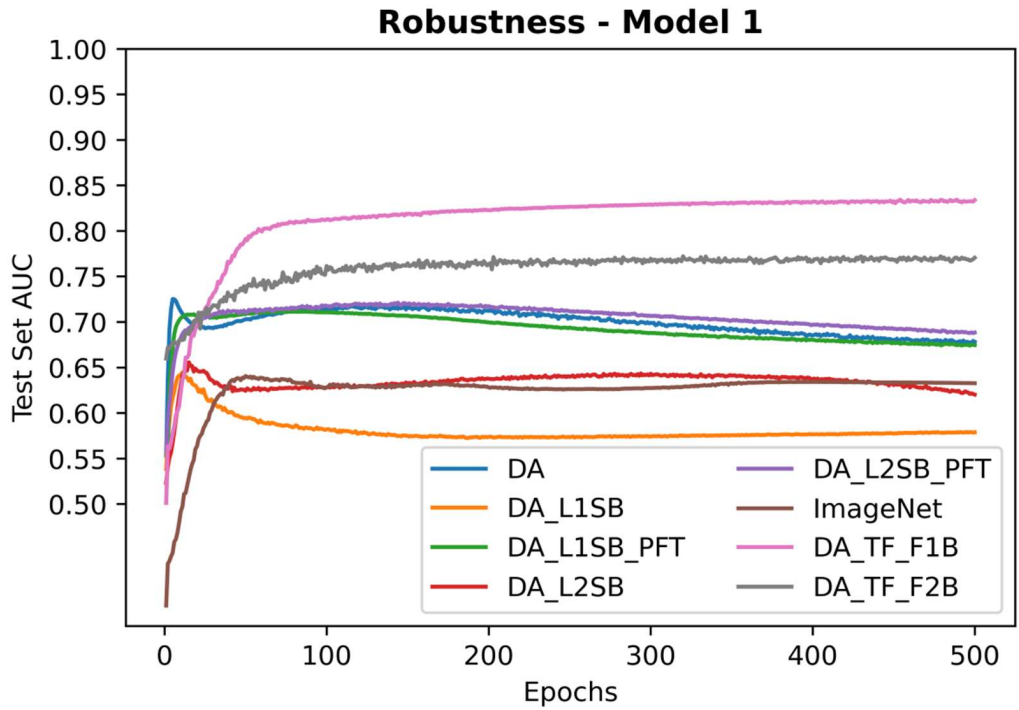

**Fig. 9.** AUC for brain tumor type classification on the external dataset (Kaggle BTTC) for model 1

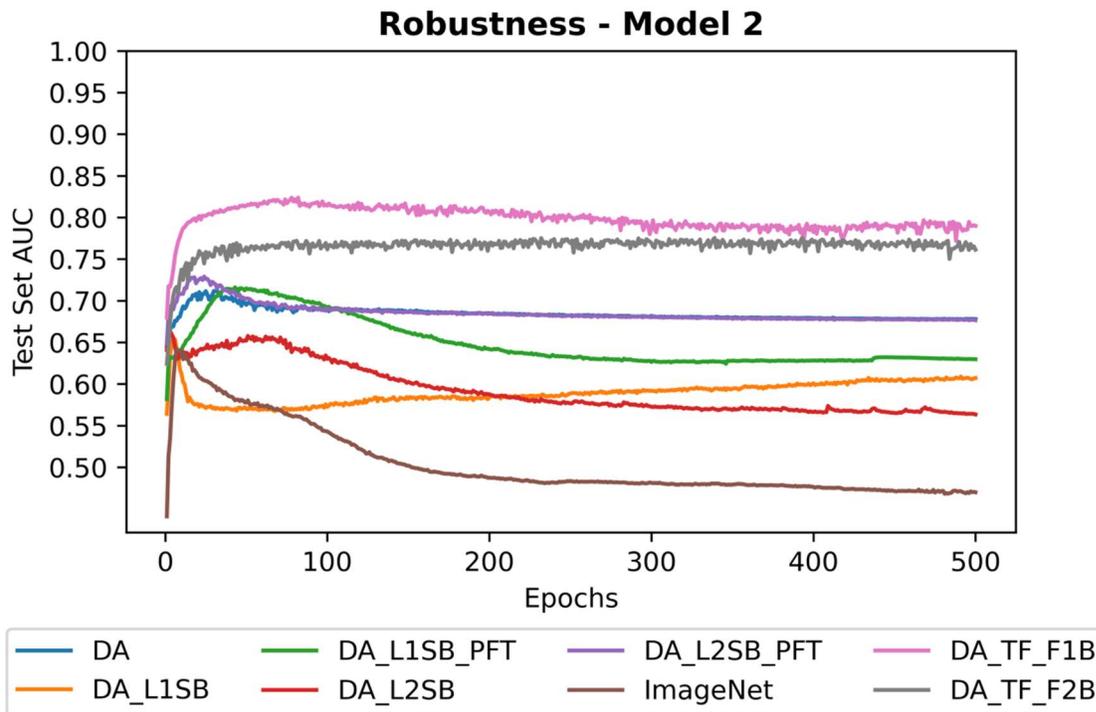

**Fig. 10.** AUC for brain tumor type classification on the external dataset (Kaggle BTTC) for model 2

**5.4 Main Findings**

To gain insight into the robustness of all the strategies, Fig. 11 and 12 (for models 1 and 2, respectively) show the performance dip from the development dataset to the external dataset using box plots for each strategy. Because of many outliers in the box plot, the default value of 1.5 as multiplier for the whiskers was replaced with fixed quantiles of 1% and 99%.

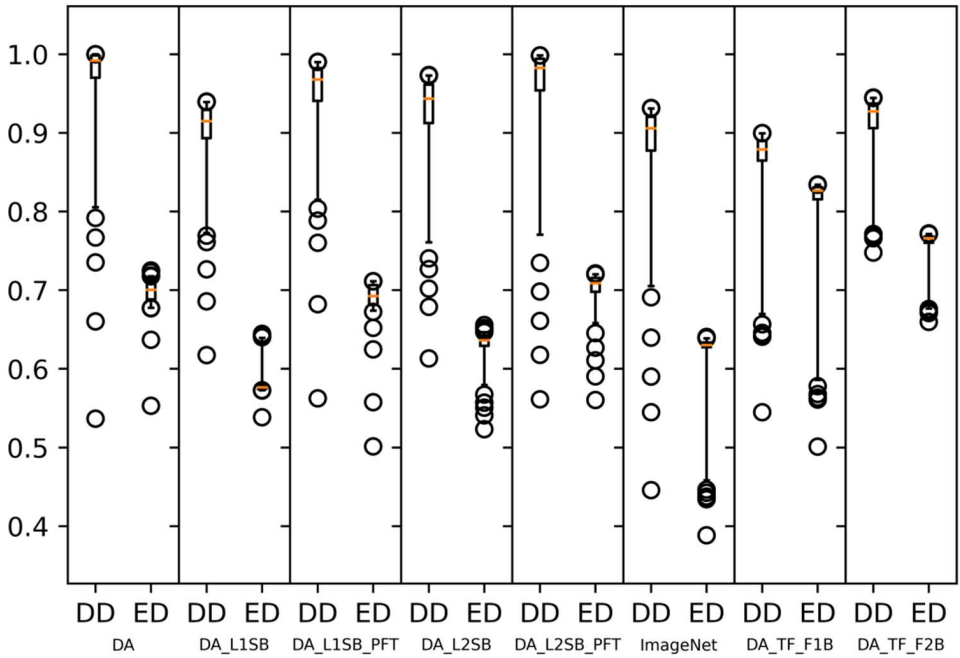

**Fig. 11.** The relative performance of strategies on the development and external datasets (labeled as DD and ED respectively on the x-axis) using the grouped box plots for model 1

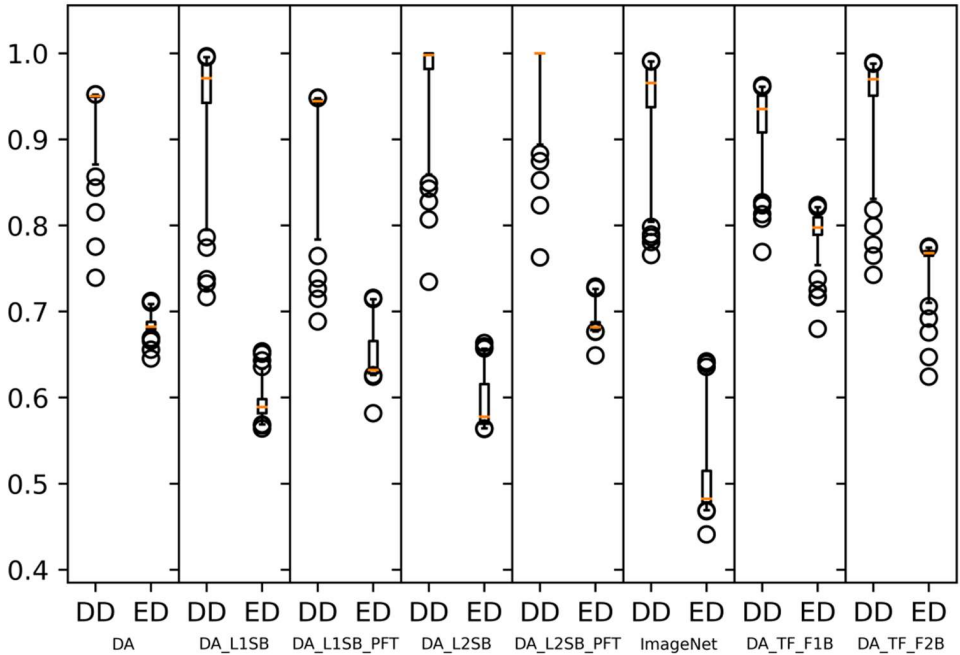

**Fig. 12.** The relative performance of strategies on the development and external datasets (labeled as DD and ED respectively on the x-axis) using the grouped box plots for model 2

Following are some of the findings/trends that can be inferred from the results of all the experiments:

- Partial DAPT, when performed on a sufficient number of layers (DA_L2SB), performs better on downstream tasks than the full DAPT strategy (DA), as can be seen from Fig. 5, 6, and 8. This finding aligns with the study's (Tajbakhsh et al., 2016) finding that fine-tuning sufficient layers outperform training all the layers using the downstream data. However, the representations learned by the full DAPT strategy generalize better to the external dataset (Fig. 9 and 10). This trend can be seen clearly from Fig. 11 and 12, where the performance dip for DA_L2SB is greater than DA for models 1 and 2.

- The hybrid strategy of progressive fine-tuning (DA_L2SB_PFT) achieved better results than the DA strategy on the development and external datasets, as shown in Fig. 5-10. There is an almost similar performance dip (from development to the external dataset) for DA_L2SB_PFT and DA for model 1 and slightly lower for DA in model 2 (Fig. 11 and 12). However, the overall results of DA_L2SB_PFT are slightly better both on development and external datasets.

- The proposed hybrid DAPT strategies outperformed their partial DAPT counterparts in all the downstream experiments (development and external datasets). Although progressive fine-tuning is expensive in terms of resources compared to partial DAPT, hybrid DAPT strategies outperform the corresponding partial DAPT strategies in accuracy and robustness. The performance dip of the hybrid DAPT strategies (as can be seen from Fig. 11 and 12) as compared to their partial DAPT counterparts is lesser for models 1 and 2.

- Simplified versions of standard architectures are the best in terms of robustness and computational cost, as can be seen from Fig. 9 and 10 while achieving modest results on the development datasets (Fig. 5-8). Similarly, it can be seen from Fig. 11 and 12 that these architectures have a minimum performance dip from the development to the external dataset. Fig. 13 and 14 also confirm that the same simplified architectures, when evaluated without using DAPT perform poorly on the development and external dataset. Therefore, the robustness of these architectures should be attributed to their simplicity as well as the DAPT performed on top of ImageNet weights.

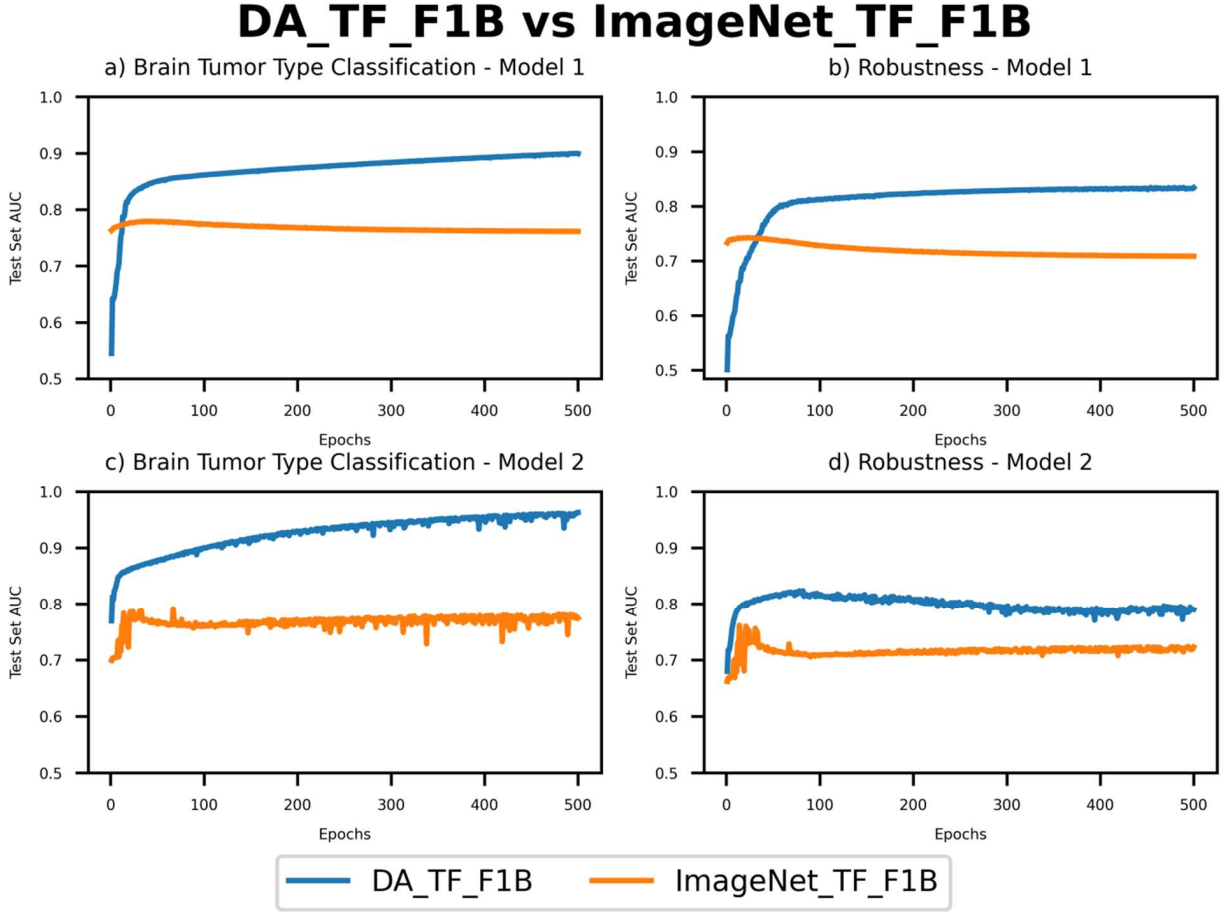

**Fig. 13.** Performance comparison of the simplified architecture based on the first one block of ResNet50 (DA_TF_F1B versus ImageNet_TF_F1B)

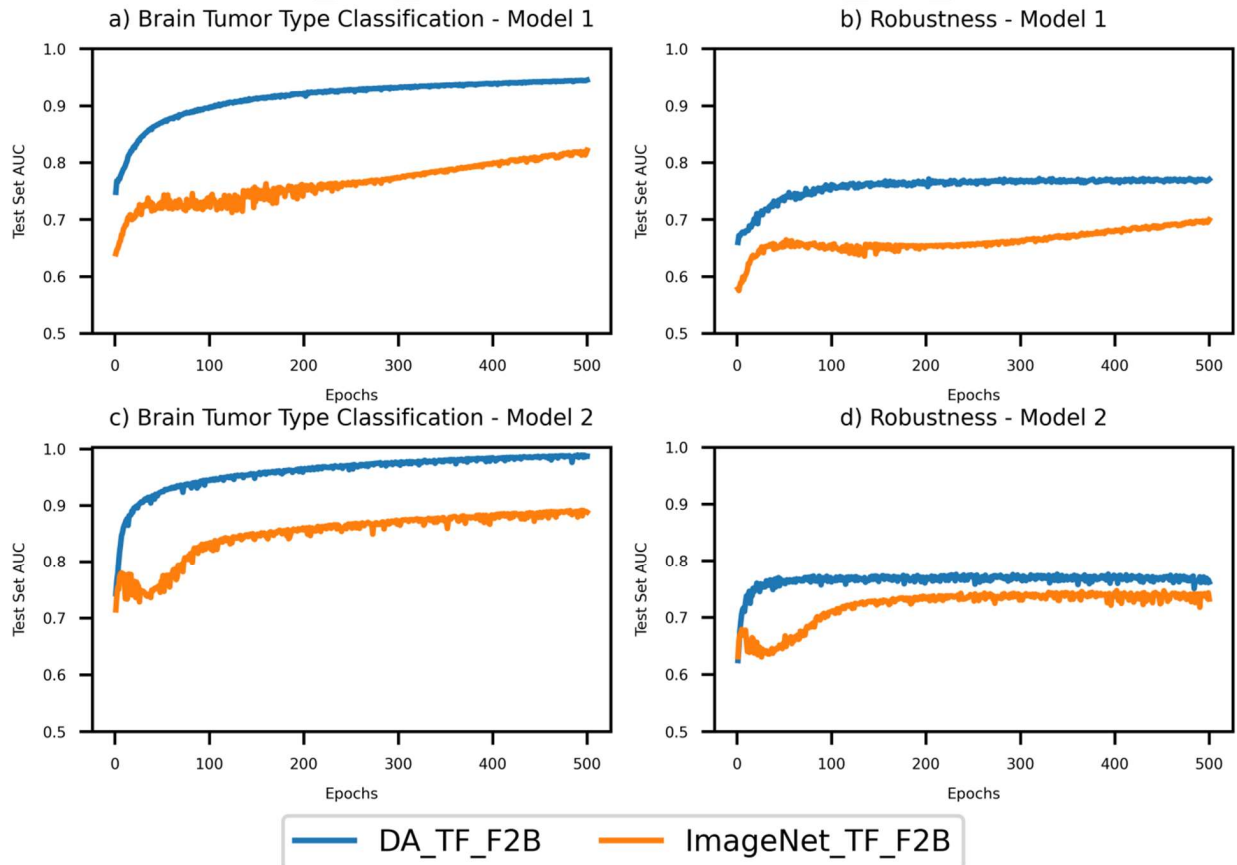

**Fig. 14.** Performance comparison of the simplified architecture based on the first two blocks of ResNet50 (DA_TF_F2B versus ImageNet_TF_F2B)

- Comparison of the performance of ImageNet pre-training on ResNet50 (ImageNet strategy) and its simpler variants (ImageNet_TF_F1B and ImageNet_TF_F2B) reveals that simpler architectures have better robustness even with pre-training using out-of-domain data, as can be seen in Fig. 15. This poor performance of the complete architecture is more pronounced when using a relatively complex downstream classifier (model 2). However, the complete architecture performed better on the development dataset.

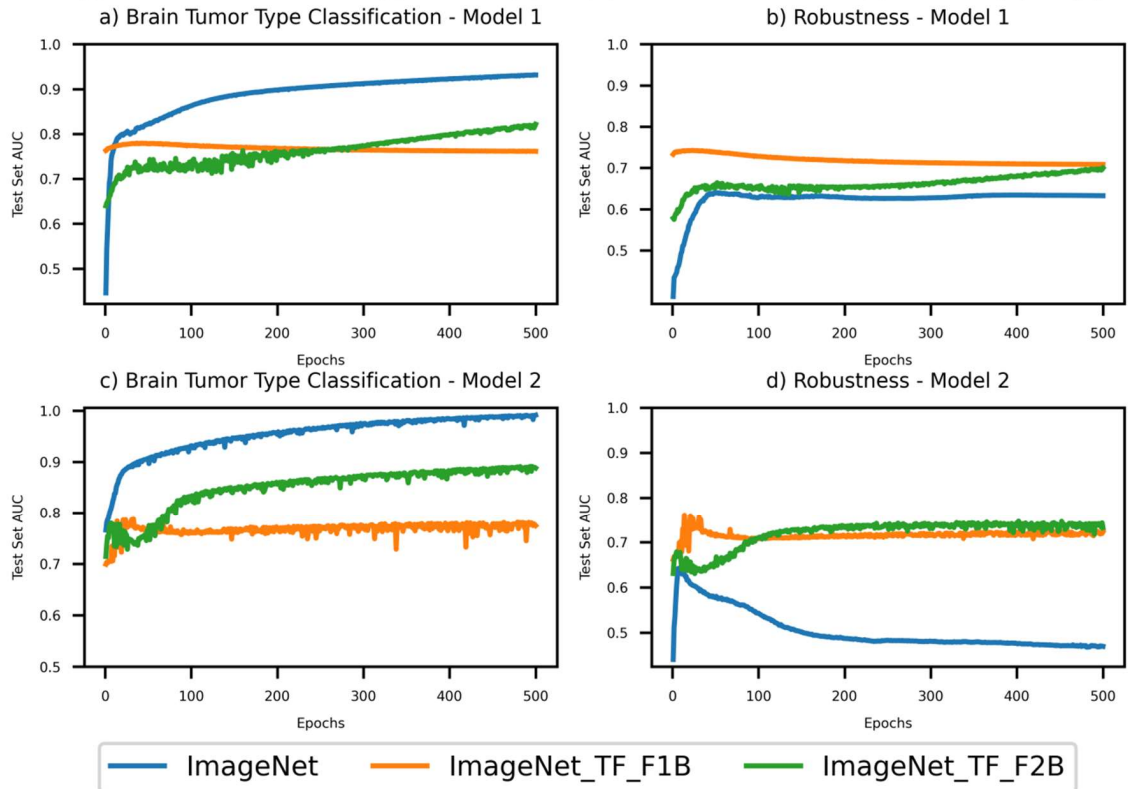

**Fig. 15.** Performance comparison of ImageNet pre-trained architectures using complete ResNet50 vs. its first one block and first two blocks (ImageNet versus ImageNet_TF_F2B versus ImageNet_TF_F1B) on development and external datasets

## 6. Conclusion and Future Works

The study proposed DAPT strategies that are computationally efficient and bridge the domain difference between the pre-training and the downstream datasets better than the full DAPT technique. Representations learned through various strategies were used in downstream classification tasks, and performance was used to gauge the quality of the representations learned. The results showed that hybrid DAPT techniques outperformed the partial and full DAPT methods on the development and external datasets. Although the simpler architectures achieved moderate performance on the development dataset but performed best on the external cohort.

Although the achieved results are encouraging, more studies are needed to repeat the experiments on other domains and other architectures. The benefits will be more pronounced if the proposed methods generalize well across other domains and complex architectures. Moreover, the simpler architectures will be more suitable for edge devices. A future direction worth exploring is to use self-supervised DAPT on simplified architectures. As self-supervised pre-training has surpassed the supervised pre-training (Azizi et al., 2021; Hosseinzadeh Taher et al., 2021), it will be interesting to see how self-supervised methods perform on the development and external dataset with these simplified architectures using DAPT.

Moreover, the study (Raghu et al., 2019) has shown that initializing the weights with a Gabor filter achieves much better results than the random initialization of weights and comparable results to the ImageNet pre-trained weights. It is, therefore, needed to evaluate the performance of various strategies when initialized with handcrafted filters to save the computational cost on generic dataset pre-training. Finally, the study (Diffenderfer et al., 2021) has shown that different techniques for simplifying neural network architectures may be combined to achieve even better results. Therefore, the performance of the proposed DAPT strategies in combination with the lottery-ticket-based methods would be worth exploring.

## Declaration of Competing Interest

The authors declare that they have no known competing financial interests or personal relationships that could have appeared to influence the work reported in this paper.

## Appendix A. Base Architecture (ResNet50)

Fig. A1 shows the overall structure of ResNet50 architecture used in the study and the annotations to highlight the portions of the architecture whose layers were made trainable or used in the simpler variants.

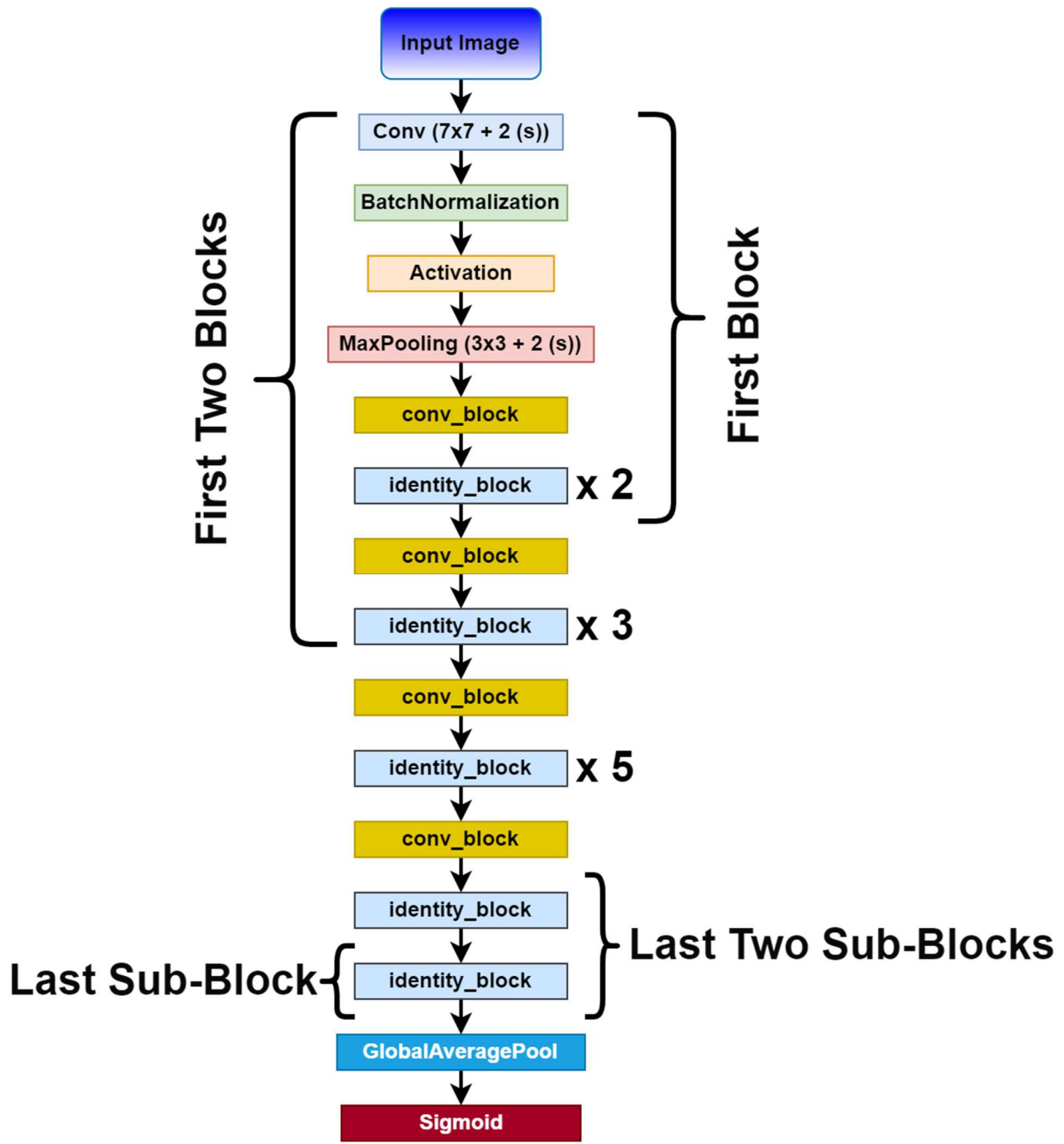

**Fig. A1.** ResNet50 architecture annotated to highlight the portions used in different DAPT strategies (Ji, Huang, He, & Sun, 2019)

Fig. A2 shows the inside structure of the conv_block and identity_block within the ResNet50 architecture.

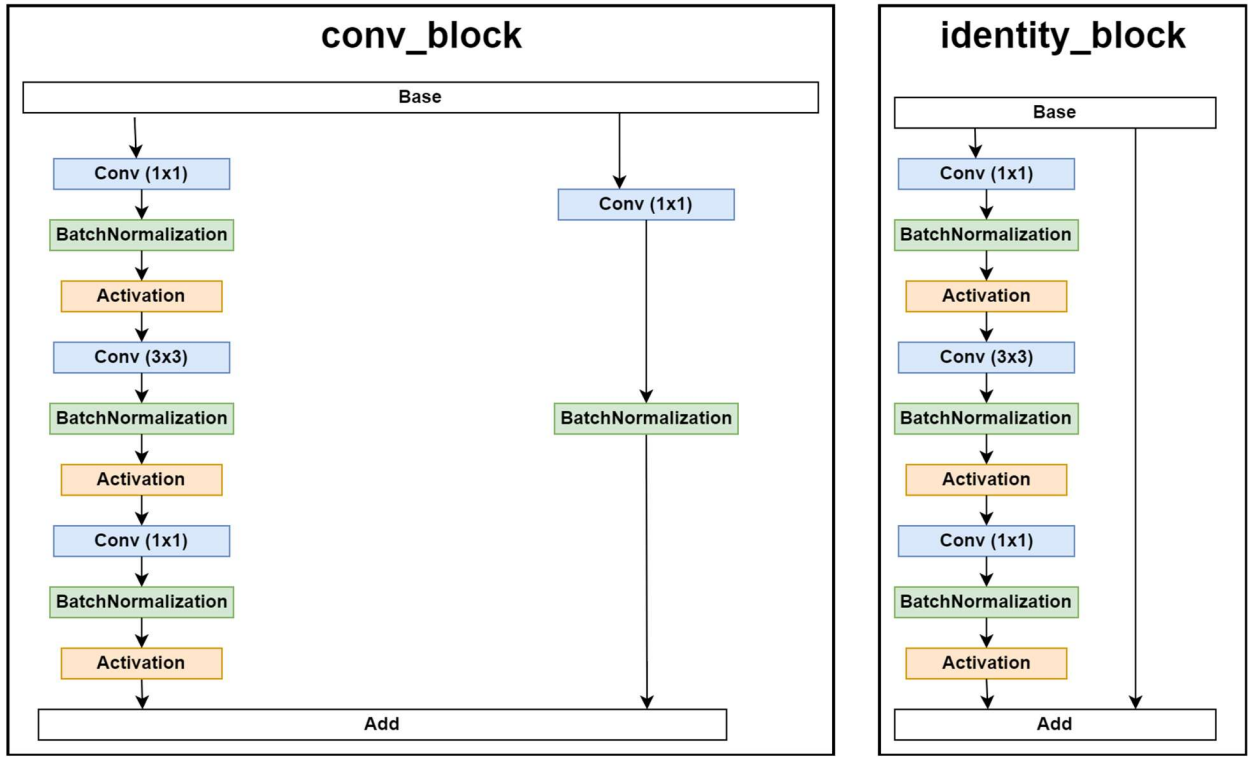

**Fig. A2.** Structure of the conv_block and identity_block within ResNet50 architecture (Ji et al., 2019)

## Appendix B. Computational Cost

To calculate the energy consumption of different DAPT strategies, the free tool (Lannelongue et al., 2021) was used. The tool calculates the energy consumption based on the running time of the algorithm along with the hardware settings (number of GPUs used, cloud vs. the local server, location of the cloud server, etc.). The first step was to calculate the total running time for each strategy. Table B1 shows each strategy's phase-wise and total running time and energy consumption in kWh. Other settings used for each strategy are given in Table B2. As all the experiments were conducted using the same hardware platform, the difference in energy consumption of these strategies is only because of their running time.

**Table B1.** Running time (phase-wise and total) and energy consumption for each DAPT strategy

| Strategy Name | Phase | Epochs | Phase-Wise Running Time (Hours:Minutes) | Total Running Time (Hours:Minutes) | Energy Consumption (kWh) |
|---|---|---|---|---|---|
| ImageNet | - | | 0 | | |
| ImageNet_F1B | - | | 0 | | |
| ImageNet_F2B | - | | 0 | | |
| DA_L1SB | 1 | 1000 | 132:12.163 | 154:37.967 | 67.93 |
| | 2 | | 22:25.804 | | |
| DA_L2SB | 1 | | 132:12.163 | 159:39.529 | 70.12 |
| | 2 | | 27:27.366 | | |
| DA | 1 | | 132:12.163 | 215:56.988 | 94.86 |
| | 2 | | 83:44.825 | | |
| DA_TF_F1B | 1 | | 40:14.181 | 69:14.181 | 30.41 |
| | 2 | | 29:00 | | |

| | | | | |
|---|---|---|---|---|
| DA_TF_F2B | 1 | 68:17.844 | 128:7.572 | 56.28 |
| | 2 | 59:49.728 | | |
| DA_L1SB_PFT | 1 | 132:12.163 | 175:4.021 | 76.90 |
| | 2 | 14:56.916 | | |
| | 3 | 27:54.942 | | |
| DA_L2SB_PFT | 1 | 132:12.163 | 179:1.836 | 78.64 |
| | 2 | 18:54.731 | | |
| | 3 | 27:54.942 | | |

**Table B2.** Parameters for Green Algorithms tool (Lannelongue et al., 2021)

| Parameter | Value |
|---|---|
| Type of cores | GPU |
| Number of GPUs | 1 |
| Model | TPU V2 |
| Memory available (in GB) | 35 |
| Select the platform used for the computations | Cloud computing |
| | Other |
| Select location | North America |
| | United States of America |
| | Any |
| Do you know the real usage factor of your GPU? | No |
| Do you know the Power Usage Efficiency (PUE) of your local data centre? | No |
| Do you want to use a Pragmatic Scaling Factor? | No |

The performance of different strategies in relation to their computational cost (memory in terms of parameters and energy consumption in kWh) has been visualized in Fig. B1.

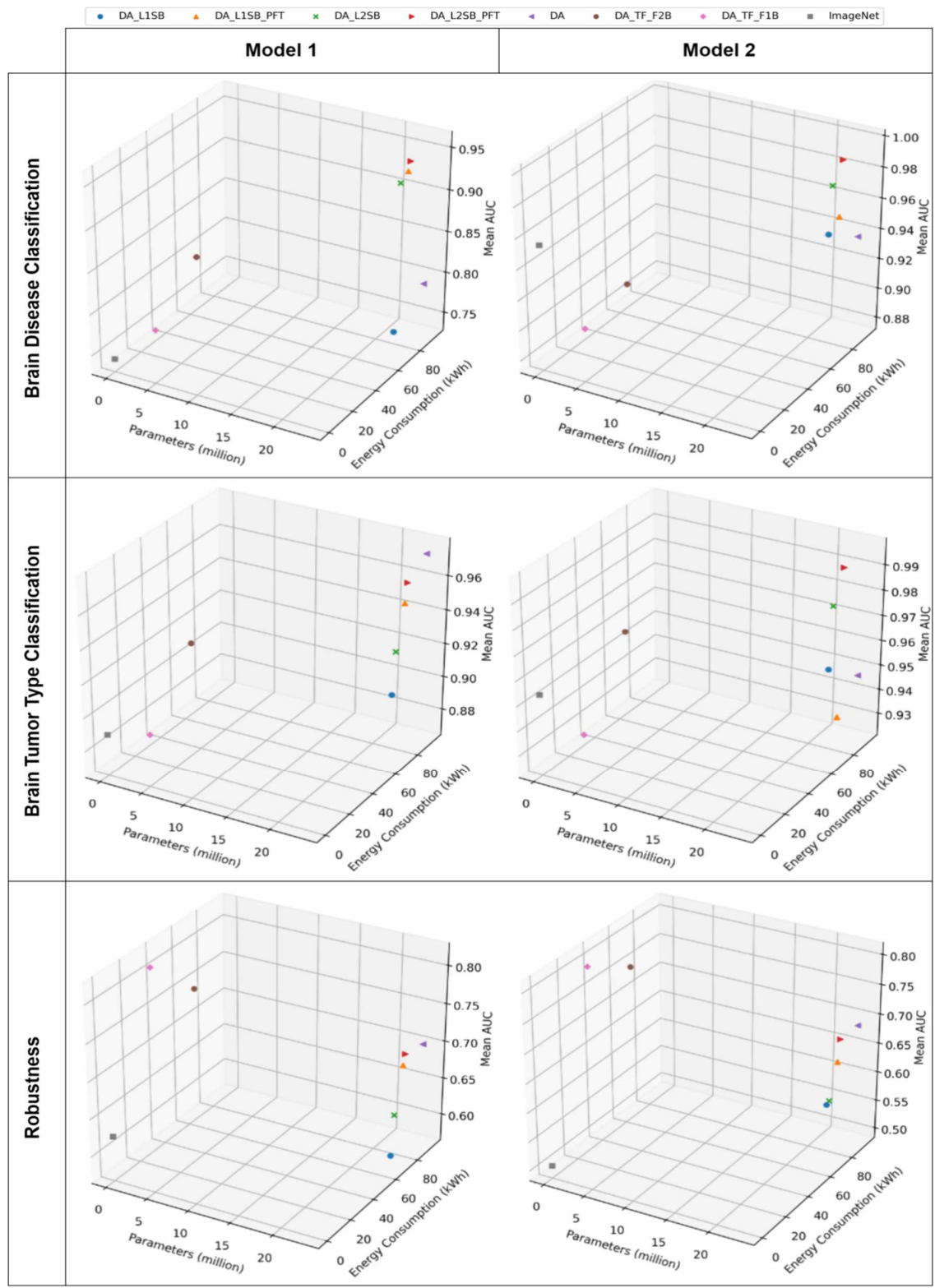

**Fig. B1.** Performance comparison of different strategies in relation to their memory usage (parameters) and energy consumption